\documentclass[a4paper,11pt]{article}
\pdfoutput=1 

\usepackage{jcappub} 

\usepackage[T1]{fontenc} 

\usepackage{graphicx}
\usepackage{ulem}

\newcommand{\Hinf}{{H_{\infty}}}
\newcommand{\kms}{{\rm\,km\,s^{-1}}}
\newcommand{\kmsmpc}{\kms{\rm\,Mpc^{-1}}}
\newcommand{\etal}{{\it et al.}}

\title{\boldmath Observational constraints on conformal time symmetry,
  missing matter and double dark energy}

\author[a,b,c]{J. Alberto V\'azquez}
\author[d,e]{S. Hee}
\author[e,1]{M.P. Hobson\note{Corresponding author.}}
\author[d,e]{A.N. Lasenby}
\author[f]{M. Ibison}
\author[d,e]{M. Bridges}

\affiliation[a]{Brookhaven National Laboratory, \\2 Center Road,
   Upton, NY 11973, USA}
\affiliation[b]{C\'atedras CONACYT, Departamento de Fisica, 
Centro de Investigaci\'on y de Estudios Avanzados del IPN, \\
A.P. 14-740, 07000 Mexico, D.F., Mexico}
\affiliation[c]{Instituto de Ciencias Fsicas, Universidad Nacional Autonoma de
Mexico, \\Apdo. Postal 48-3, 62251 Cuernavaca, Morelos, Mexico}
\affiliation[d]{Kavli Institute for Cosmology, \\Madingley Road, 
Cambridge CB3 0HA, UK}
\affiliation[e]{Astrophysics Group, Cavendish Laboratory, \\JJ Thomson
  Avenue, Cambridge CB3 0HE, UK}
\affiliation[f]{Institute for Advanced Studies at Austin, \\11855
  Research Boulevard, Austin, TX 78759, USA}

\emailAdd{javazquez@icf.unam.mx}
\emailAdd{sh767@cam.ac.uk}
\emailAdd{mph@mrao.cam.ac.uk}
\emailAdd{anthony@mrao.cam.ac.uk}
\emailAdd{ibison@ias-austin.org}
\emailAdd{michael.bridges@gmail.com}

\abstract{The current concordance model of cosmology is dominated by two
mysterious ingredients: dark matter and dark energy.  In this paper,
we explore the possibility that, in fact, there exist two dark-energy
components: the cosmological constant $\Lambda$, with
equation-of-state parameter $w_\Lambda=-1$, and a `missing matter'
component $X$ with $w_X=-2/3$, which we introduce here to allow the
evolution of the universal scale factor as a function of conformal
time to exhibit a symmetry that relates the big bang to the future
conformal singularity, such as in Penrose's conformal cyclic
cosmology.  Using recent cosmological observations, we constrain the
present-day energy density of missing matter to be
$\Omega_{X,0}=-0.034 \pm 0.075$. This is consistent with the standard
$\Lambda$CDM model, but constraints on the energy densities of all the
components are considerably broadened by the introduction of missing
matter; significant relative probability exists even for $\Omega_{X,0}
\sim 0.1$, and so the presence of a missing matter component cannot be
ruled out. As a result, a Bayesian model selection analysis only slightly
disfavours its introduction by 1.1 log-units of
evidence. Foregoing our symmetry requirement on the conformal time
evolution of the universe, we extend our analysis by allowing $w_X$ to
be a free parameter. For this more generic `double dark energy' model,
we find $w_X = -1.01 \pm 0.16$ and $\Omega_{X,0} = -0.10 \pm 0.56$,
which is again consistent with the standard $\Lambda$CDM model,
although once more the posterior distributions are sufficiently broad
that the existence of a second dark-energy component cannot be ruled
out.  The model including the second dark energy component also has an
equivalent Bayesian evidence to $\Lambda$CDM, within the estimation
error, and is indistinguishable according to the Jeffreys
guideline.}


\begin{document}
\maketitle
\flushbottom


\section{Introduction}

Over the past two decades, cosmological observations have confirmed that
the background expansion of the universe is accelerating
\cite{Riess98, Perlmutter99}. This remarkable phenomenon is usually
explained by assuming the existence of a single dark-energy component,
often modelled as a perfect fluid with a (generally time-dependent)
equation-of-state parameter $w(z)$ that results in it exhibiting a
negative pressure. The simplest form of dark energy is a cosmological
constant $\Lambda$, which corresponds to a \textit{constant} equation
of state $w_{\Lambda}=-1$. Together with cold dark matter, which is
key to explaining the evolution of structure in the universe, the
cosmological constant gives rise to the standard $\Lambda$CDM model,
which provides a good fit to existing cosmological observations.
Nonetheless, there have been a large number of other exotic forms of
matter proposed to provide alternative explanations for the current
accelerating universal expansion \cite{Copeland06,Durrer08},
including, for example, topological defects \cite{Vilenkin85}.

In this paper, we remain focussed on the $\Lambda$CDM model, but with
the inclusion of a second, additional, dark energy component, having a
different equation of state parameter. One of the motivations for
exploring such a possibility arises from Penrose's `conformal
cyclic cosmology' (CCC) model \cite{Penrose}, which posits a cyclic
universe in which the ultimate infinitely expanded state of one phase
(or `aeon') is identified with the initial singularity of the next.
One way of realising such a model is to relate the future conformal
singularity to the big bang, which leads one to investigate the
symmetries of the Friedmann equations when written in terms of
conformal time. Interestingly, as we will show, one finds that if the
evolution of the universal scale factor $a$ is to have an appropriate
symmetry in conformal time, one requires the existence of an
additional component with equation-of-state $w=-\tfrac{2}{3}$.

Indeed, even without the above considerations, the standard form of
the Friedmann equation written in terms of cosmic time hints at such a
hitherto neglected additional component.  For a homogeneous and
isotropic universe described by the Friedmann--Robertson--Walker (FRW)
metric, the Friedmann equation describing
the dynamical evolution of the scale factor $a(t)$ can be written
as\footnote{It is useful for later purposes to adopt the convention
  that the subscript $_0$ refers to evaluation at the time $t_0$ at
  which $a(t)=1$, but that there is no necessary link with the
  present-day; $t_0$ is merely some reference or `fiducial' time.}
\begin{equation}\label{eq:Fried}
   \left( \frac{H}{H_0} \right)^2 = \sum_i \Omega_{i,0} \, a^{-3 (1+w_i)},
  \end{equation}
\noindent
where $H=\dot{a}/a$ is the Hubble parameter (the dot denotes
differentiation with respect to cosmic time $t$), and the energy
density $\rho_i$ of each of the constituent components of the universe
is taken into account through a corresponding density
parameter $\Omega_{i,0} = 8\pi G \rho_{i,0}/(3H_0^2)$.   The equation-of-state parameters are $w_i$, which we will assume throughout to
be time-independent. The summation in (\ref{eq:Fried}) also includes
the curvature density parameter $\Omega_{k,0}$, so that $\sum_i
\Omega_{i,0}=1$.

In the $\Lambda$CDM model, the total density parameter is usually
taken to comprise of contributions from radiation ($w=\frac{1}{3}$),
matter (typically modelled as dust with $w=0$), curvature
($w=-\frac{1}{3}$), and the cosmological constant ($w=-1$).  These are
listed in Table~\ref{tab:omega}, in which one can see an obvious `gap'
that we term `missing matter' with $w=-\frac{2}{3}$.  Interestingly,
forms of matter have been proposed for which $w=-\frac{2}{3}$, such as
domain walls \cite{Conversi04, Battye99,Mithani12}, or particular
scalar field models \cite{Caldwell98}.
\begin{table}
\begin {center}
\begin{tabular}{rcl}
\hline
\vspace{0.1cm}
$w_i$           \qquad \qquad   & {\rm component}               & \qquad    $\Omega_i$          \\
\hline
\vspace{0.1cm}	
$1/3$           \qquad  \qquad  & {\rm radiation}               & \qquad    $\Omega_{\rm r}$    \\
\vspace{0.1cm}	
$0$             \qquad  \qquad  & {\rm matter (dust)}           & \qquad    $\Omega_{\rm m}$    \\
\vspace{0.1cm}	
$-1/3$          \qquad  \qquad  & {\rm curvature}               & \qquad    $\Omega_{k}$        \\
\vspace{0.1cm}	
$-2/3$          \qquad  \qquad  & {\rm missing matter ?}        & \qquad    $\Omega_{X}$        \\
\vspace{0.1cm}	
$-1$            \qquad  \qquad  & {\rm cosmological constant}   & \qquad    $\Omega_{\Lambda}$  \\
\hline
\end{tabular}
\caption{Canonical equation-of-state parameters for different constituents of the
  universe.}
\label{tab:omega}
\end{center}
\end{table}
It should be noted, of course, that the true equation-of-state
parameters for matter and radiation will, in general, differ from
the canonical values listed in Table~\ref{tab:omega} (although these
values are assumed in most cosmological analyses). For example,
non-relativistic matter does not have exactly zero pressure ($w=0$),
but a pressure proportional to $(v/c)^2$. Similarly, relativistic
particles such as massive neutrinos have an equation-of-state
parameter slightly less than $w=\frac{1}{3}$, which changes with
cosmic epoch. Nonetheless, these deviations from the canonical values
are small and the equation-of-state parameters for curvature and a
pure cosmological constant are fixed to the values listed in
Table~\ref{tab:omega}. Hence the suggestion of a missing component
remains a distinguishable (distinct) possibility.

Once one admits the possibility of adding an extra component, however,
it is natural to extend one's investigation by allowing its
equation-of-state parameter to vary, rather than fixing it to
$w=-\frac{2}{3}$. This more generic `double dark energy' model comes
at the cost of breaking the desired symmetry of the Friedmann equation
in conformal time, and hence loses contact with Penrose's `Cycles of
Time' proposal. Nonetheless, such a model is also of interest
in its own right since the observed acceleration of the universal
expansion may be driven by more than just a single dark-energy
component. We note that a generic two-component model of dark energy
has previously been considered in \cite{Gong07}.

The structure of this paper is as follows. In
Section~\ref{sect:rad-only}, we begin by considering the symmetry of
the evolution of the scale factor $a$ as a function of conformal time
for the simplified case of a spatially-flat, radiation-filled universe
with a cosmological constant, and then pass to the more general case
with matter and curvature included in Section~\ref{sect:matt-inc}.  We
give a brief summary in Section~\ref{sec:phenom} of the phenomenology
of an additional missing matter component with $w=-\frac{2}{3}$ by
investigating its effect on the expansion history of the universe, in
particular the distance-redshift relation, and on the evolution of
perturbations, through the cosmic microwave background (CMB) and
matter power spectra. In Section~\ref{sec:analysis}, we describe our
Bayesian parameter estimation and model selection analysis methodology
and the cosmological data sets used to set constraints on our `missing
matter' and `double dark energy' models. The results of these analyses
are given in Section~\ref{sec:results} and our conclusions are
presented in Section~\ref{sec:conclusions}.


\section{Conformal time development of a radiation-filled
flat-$\Lambda$ universe}

\label{sect:rad-only}

We begin by considering the evolution of the scale factor in a
spatially-flat, radiation-filled universe with a cosmological
constant. Such a model may seem rather artificial at first, but in
fact corresponds well to the initial and final stages of a real
universe containing matter, since radiation dominates at the beginning
and $\Lambda$ dominates at the end. Indeed, as argued by Penrose in
the CCC model, at the two extremes of the big bang and future
conformal singularity, only massless particles are likely to be
present.

The time development of the main parameters of such a universe can be
expressed most simply in terms of cosmic time $t$. Using the definition
$H^2_\infty\equiv\Lambda/3$ (and setting $c=1$ throughout), one finds
\begin{equation}
\begin{aligned}
a(t)            &= a_{\rm eq} \sinh^{1/2} (2 \Hinf t),\\
\rho_{\rm r}(t) &= \frac{\rho_{\rm r,0}}{a^4(t)},\\
H(t)            &= \Hinf \coth(2 \Hinf t),
\end{aligned}
\label{eqn:t-sols}
\end{equation}
where $a_{\rm eq}^4 = 8 \pi G \rho_{\rm r,0}/(3H^2_\infty)$ and the
subscript $_{\rm eq}$ refers to the instant $t_{\rm eq}$ at which the
radiation energy density $\rho_{\rm r}$ is equal to the vacuum energy
density $\Lambda/(8\pi G)$, and the subscript $_0$ refers to the time
$t_0$ when $a=1$, as mentioned above.

One may also write these solutions in terms of conformal time $\eta$,
related to cosmic time by $d\eta = dt/a$.  Indeed, as discussed in
\cite{Ibison11}, a major motivation for working in terms of $\eta$ is
that, for currently accepted values of the density parameters
$\Omega_{i,0}$, the conformal time intervals since the Big Bang
($a=0$) and until the conformal singularity ($a=+\infty$) are both
finite. By contrast, although the cosmic time since the Big Bang is
finite, the future singularity occurs at $t=\infty$. This asymmetry
means that it is more natural to work in terms of conformal time, if
one is to realise scenarios such as the CCC model. It is worth noting
that, like cosmic time, which corresponds to the proper time of
comoving observers, conformal time also has a clear operational
definition as the time kept by a (Marzke--Wheeler) clock whose `tick'
is the bounce of a light pulse confined to a pair of parallel mirrors
moving, and therefore separating, with the Hubble flow
\cite{Marzke64}.

\begin{figure}[tbp]
\centering
\includegraphics[width=0.7\linewidth]{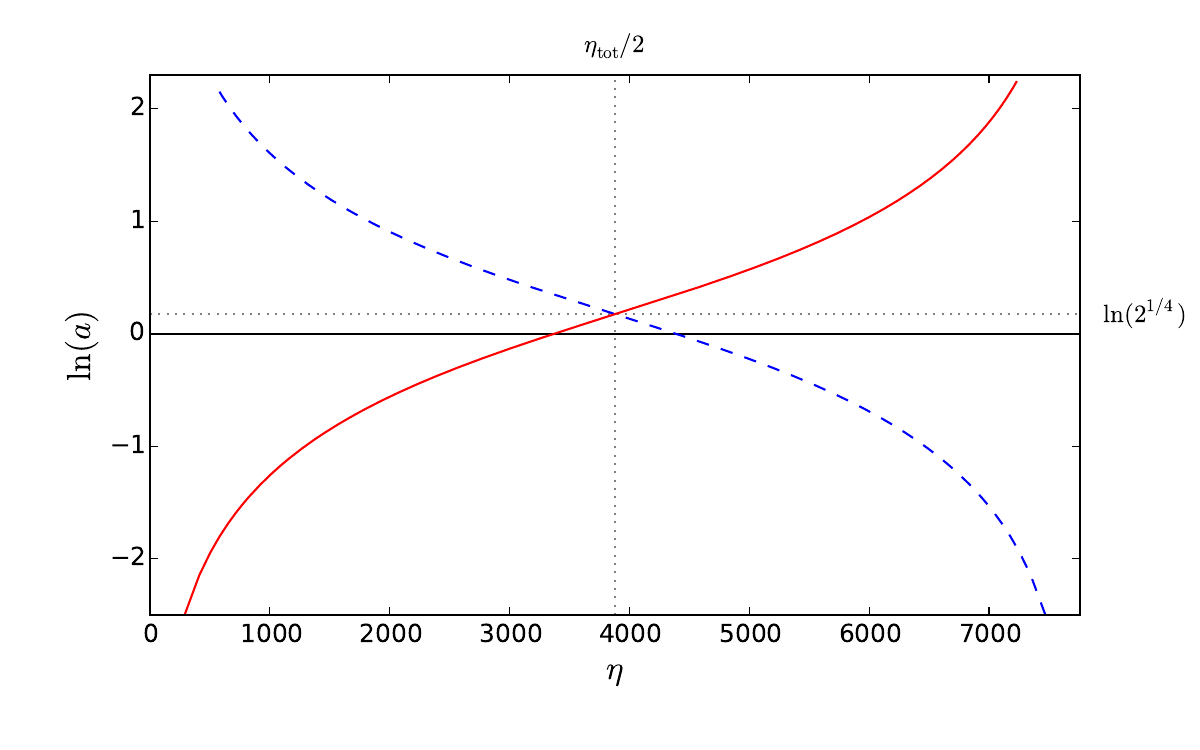}
\caption{Evolution (red solid curve) of the natural logarithm of the
  scale factor as a function of conformal time in a spatially-flat,
  radiation-filled universe with $\Lambda$ given by recent estimates
  ($\Omega_{\Lambda}=0.7$ and $H_0 = 72 \kmsmpc$), with the unit of
  time and space given by $1 {\rm \, Mpc}$. As an example, we have
  arbitrarily taken $a_{\rm eq}=2^{1/4}$. The blue dashed curve is the
  red curve reflected left-right about $\eta=\eta_{\rm tot}/2$. These
  curves are symmetrical not just left-right, but top-bottom if the
  line of reflection is taken through the value of $a$ at the
  mid-point, i.e.\ where $a=2^{1/4}$. We can thus put the curves on
  top of one another if we use the reciprocal, $\tilde{a}=a_{\rm
    eq}^2/a=\sqrt{2}/a$. Then the blue curve is flipped and slid up by
  an appropriate amount to lie on top of the red curve.}
\label{fig:simple-flip}
\end{figure}

The transition to conformal time can be carried out analytically for
the equations (\ref{eqn:t-sols}) and results in solutions expressed in
terms of elliptic functions (see Lasenby \etal, in preparation, for
further details). The important point to note here, however, is that
one may show that the `epoch of equality' $\eta_{\rm eq}$ occurs exactly
{\it half way through} the total conformal time evolution from the big
bang to the future singularity. Moreover, the evolution after equality
is {\it identical} to that before equality if one works in terms of a
reciprocal scale factor defined by $\tilde{a}=a_{\rm eq}^2/a$. This
equivalence is illustrated in Fig.~\ref{fig:simple-flip}.

Thus any radiation-filled, flat-$\Lambda$ universe has the same basic
symmetry: the development of the scale factor after the mid-point in
conformal time evolution is the reciprocal (up to an overall
multiplicative constant) of the development up to the mid-point.\footnote{In
fact, the value chosen for $a_{\rm eq}$ is arbitrary, and merely
determines the units of conformal time, once $\Lambda$ has been
specified; it is therefore sensible to use $a_{\rm eq}=1$ in this
case, so that the reciprocal relation is just $\tilde{a}=1/a$.}


\section{Inclusion of matter and curvature}
\label{sect:matt-inc}

We have just shown that for radiation-only universe with $\Lambda$ the
future conformal singularity is approached in a manner identical as a
function of $1/a$ to the way the big bang is exited as a function of
$a$. This symmetry is clearly interesting in connection with attempts,
such as the CCC model, to relate the final singularity in conformal
time to the big bang. The key question remaining is whether the
symmetry can survive the inclusion of matter and curvature. As we now
show, this is indeed the case, but only provided a suitable amount
of the component labelled ``missing matter'' in Table~\ref{tab:omega}
is present.

Making the change of variable $d\eta = dt/a$ in the Friedmann equation
(\ref{eq:Fried}) and adopting the canonical equation-of-state
parameters listed in Table~\ref{tab:omega}, including an additional
missing matter component $X$, one obtains
\begin{equation}
\frac{1}{H_0^2}\left(\frac{da}{d\eta}\right)^2 = \Omega_{{\rm
    r},0}+\Omega_{{\rm m},0}a + \Omega_{k,0}a^2 + \Omega_{X,0} a^3 +
\Omega_{\Lambda,0} a^4,
\label{eq:friedfull}
\end{equation}
where we note that the right-hand side is simply a fourth-degree
polynomial in $a$. Guided by our findings in
Section~\ref{sect:rad-only} for the radiation-only, flat-$\Lambda$
case, we make the change of variable $\tilde{a}(\eta)=\alpha^2/a(\eta)$,
where $\alpha$ is a constant. This immediately yields
\begin{equation}
\frac{1}{H_0^2}\!\!\left(\frac{d\tilde{a}}{d\eta}\right)^2
\!\!\!=  \alpha^4 \Omega_{\Lambda,0}
    + \alpha^2 \Omega_{X,0} \tilde{a}
    + \Omega_{k,0}\tilde{a}^2\\
    + \frac{\Omega_{{\rm m},0}}{\alpha^2}\tilde{a}^{3}
    + \frac{\Omega_{{\rm r},0}}{\alpha^4}\tilde{a}^{4}.
\label{eq:friedinvert}
\end{equation}
We thus obtain an identical equation in the new variable, $\tilde{a}$,
if the densities are related by
\begin{equation}
\Omega_{{\rm m},0} =\alpha^2 \Omega_{X,0}, 
\quad \text{and} \quad \Omega_{{\rm    r},0} = \alpha^4 \Omega_{\Lambda,0}.
\label{eqn:conds}
\end{equation}
Noting that the LHS of (\ref{eq:friedfull}) and (\ref{eq:friedinvert})
are invariant under $\eta\mapsto -\eta$, and the RHS of each does not
contain $\eta$ explicitly, this means that if the conditions in
equation (\ref{eqn:conds}) are satisfied, and if we measure $\eta$
from the point where $\tilde{a}=a$, i.e.\ where $a^2=\alpha^2$, then for
general $\eta$ we will have $a(\eta)a(-\eta)=\alpha^2$. The relevance of
satisfying (\ref{eqn:conds}) is that this leads to the derivatives of
$a$ and $\tilde{a}$ matching at point when $\tilde{a}=a$, which is of course
necessary if the function is to go smoothly through this point, whilst
at the same time tracing out the reciprocal behaviour. We note this
behaviour will be obtained even with curvature included, since the
the symmetry does not require any special value of $\Omega_{k,0}$.

\begin{figure}[tbp]
\centering
\includegraphics[width=0.65\linewidth]{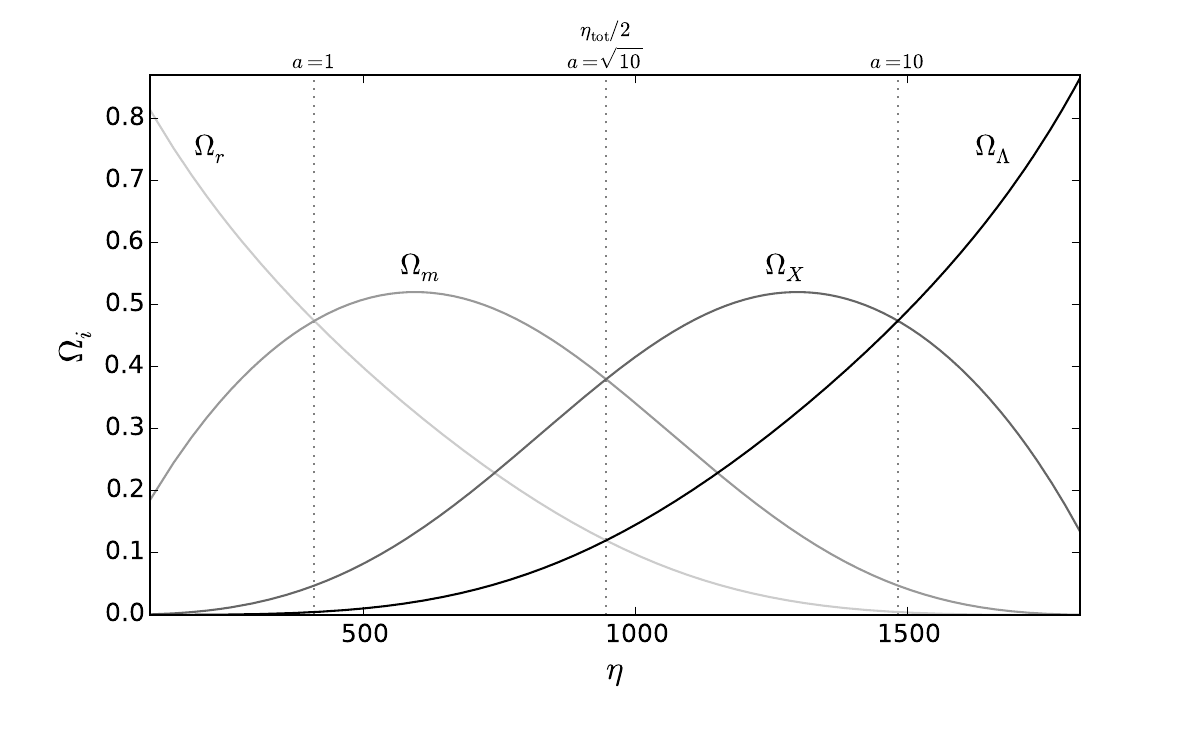}
\hfill
\includegraphics[width=0.65\linewidth]{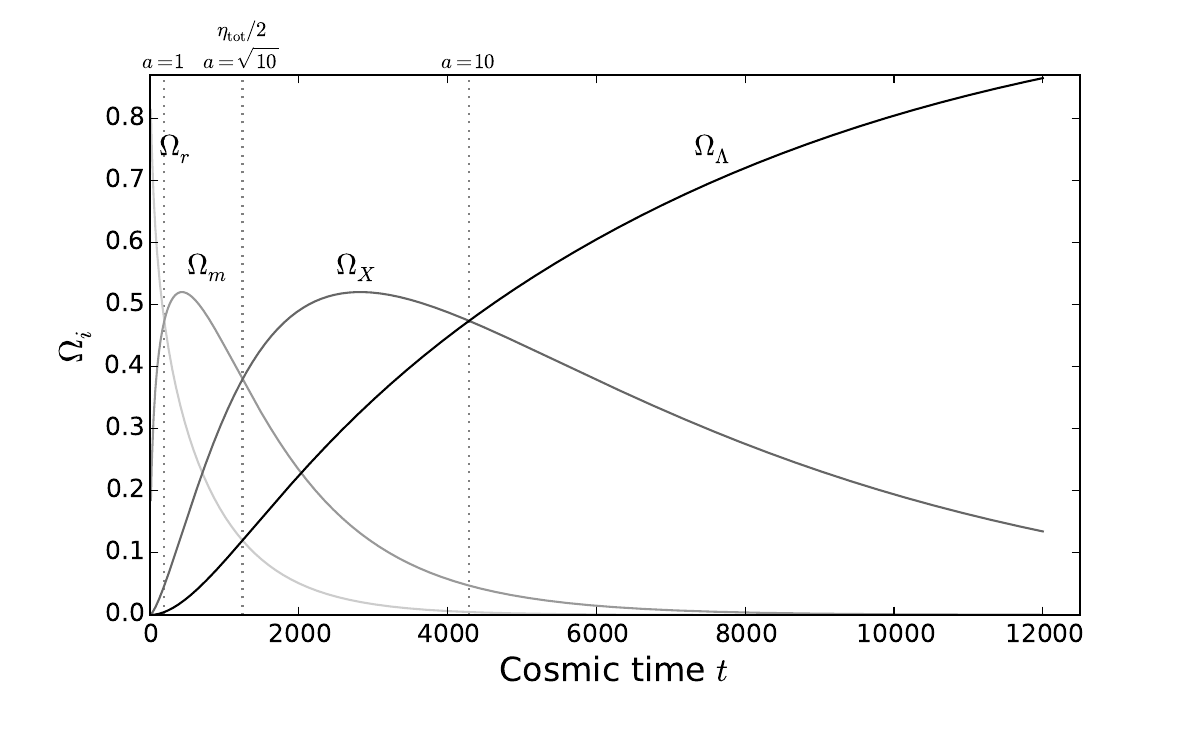}
\hfill
\includegraphics[width=0.65\linewidth]{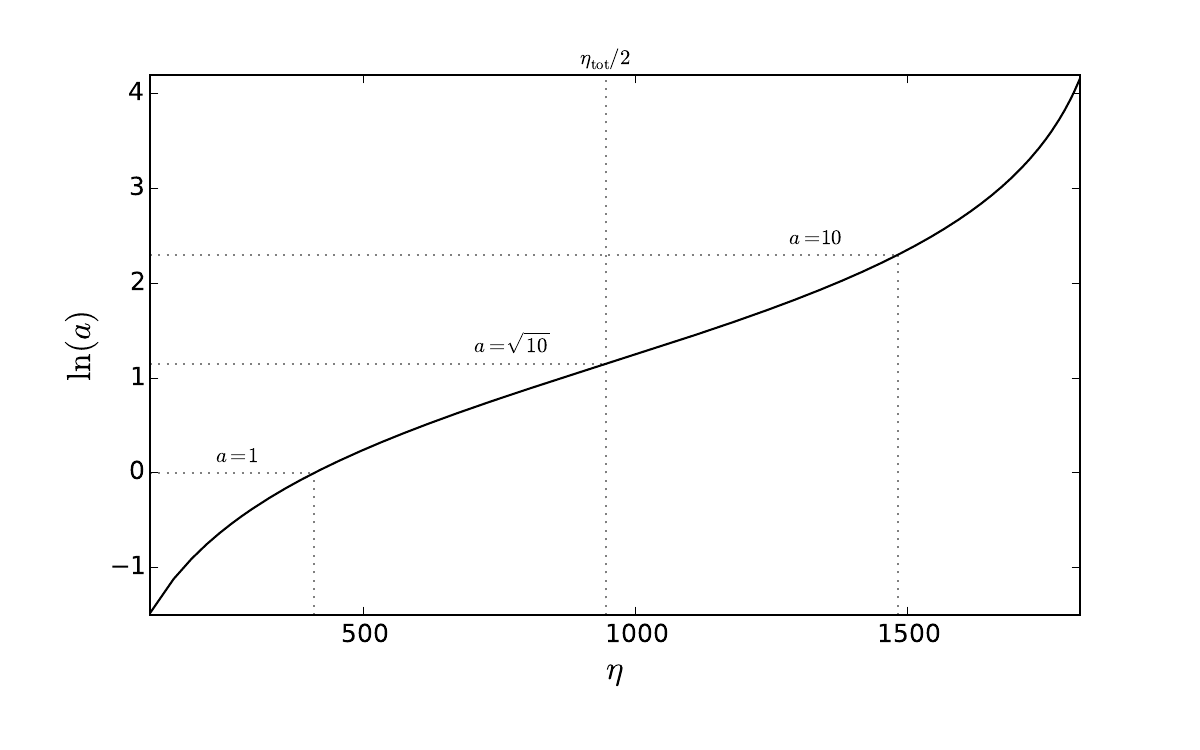}
\caption{Evolution of a spatially-flat universe with matter and
  `missing matter' present in the proportions discussed in the
  text. The top panel shows the evolution of the radiation, matter,
  missing matter and vacuum energy densities as a function of
  conformal time $\eta$, while the middle panel show the same as a
  function of cosmic time $t$. The bottom panel shows the evolution of
  the natural logarithm of the scale factor versus conformal time over
  the same period.}
\label{fig:evol-with-matter}
\end{figure}

As a concrete example of this behaviour, we show in
Fig.~\ref{fig:evol-with-matter}
the evolution of the energy densities of the components as a function
of both conformal time and cosmic time, in a spatially-flat
($\Omega_{k,0}=0$) case where equation (\ref{eqn:conds}) is satisfied,
with $\alpha^2=10$. Specifically, in this illustrative case, we have
chosen $\Omega_{{\rm m},0} = 100 \, \Omega_{\Lambda,0}$, $\Omega_{{\rm
    r},0} = 100 \, \Omega_{\Lambda,0}$ and $\Omega_{X,0} = 10 \,
\Omega_{\Lambda,0}$.  These particular values mean e.g.\ that the
radiation and matter densities should be equal at $a=1$, and the
`missing matter' and vacuum energy densities should be equal at
$a=10$, both of which can be verified easily from the bottom panel.

We see in this case that we have indeed obtained symmetry in the
density parameters about the mid-point in conformal time, and moreover
the $a(\eta)$ plot is again symmetric under flipping about the
horizontal axis going through the value at the mid-point
($a=\sqrt{10}$), meaning that it is symmetric in the inverse scale
factor in the same way as for the radiation-only case in
Section~\ref{sect:rad-only}. It is straightforward to
extend this example to include curvature, which yields the same results
as regards the symmetries.

It is worth noting that the form invariance of the dynamical laws
governing the evolution of the conformal metric scale factor to the
reciprocity transformation $\tilde{a}(\eta)=\alpha^2/a(\eta)$ implies
an indifference of the dynamics to exchange of the roles of radiation
with dark energy, and matter with `missing matter', and also to
exchange of the roles of the big bang and future conformal
singularity. Moreover, as an intrinsic symmetry of a dynamical law, this
invariance has the same status with respect to the distribution of the
various contributions to the cosmological stress-energy tensor as does
homogeneity and isotropy: it is only `broken' by
cosmological perturbations in that sense that a particular phase-space
distribution of particles in the cosmological fluid may not obey it,
but it remains valid in a statistical sense (either on large scales or
across an ensemble of universes).

As a caveat, however, one should recall that the true
equation-of-state parameters for radiation and matter (and possibly
missing matter) will, in general, differ from the canonical values
listed in Table~\ref{tab:omega} and vary with cosmic epoch, as
discussed in the Introduction.  Consequently, the RHS of
(\ref{eq:friedfull}) will not, in general, be a fourth-degree
polynomial, in which case it no longer has the opportunity to remain
form-invariant\footnote{In the case where the equation-of-state
  parameter for each component is constant, but might differ slightly
  from the canonical values listed in Table~\ref{tab:omega}, so that
  $w_i \to w_i + \frac{1}{3}\epsilon_i$, each term on the right-hand
  side of (\ref{eq:friedfull}) would be separately multiplied by the
  appropriate factor $a^{-\epsilon_i}$, whereas each term on the
  right-hand side of (\ref{eq:friedinvert}) would simply inherit the
  additional factor $\tilde{a}^{\epsilon_i}$. Thus, form-invariance
  under the reciprocity transformation would be recovered if
  $\epsilon_\Lambda = -\epsilon_{\rm r}$ and $\epsilon_X =
  -\epsilon_{\rm m}$, together with the automatic condition
  $\epsilon_k=0$.} under the reciprocity transformation
$\tilde{a}(\eta)=\alpha^2/a(\eta)$.

Nonetheless, the basic notion of symmetric behaviour at the big bang
and future conformal singularity remains valid, particularly since
at these extremes only massless particles are likely to be present (as
argued by Penrose in the CCC model). Thus, it still seems of interest
to explore the symmetry discussed here as a possible approximate
symmetry of our universe. In particular, the key to realising this
symmetry, is the existence of the `missing matter' component, which
moreover has to be present in the proportion discussed earlier, and
encoded in equation (\ref{eqn:conds}). The possibility that such a
`missing matter' component is indeed present in our universe seems
well worth testing against current cosmological observations.


\section{Phenomenology}
\label{sec:phenom}

Given the motivation presented in Sections~\ref{sect:rad-only} and
\ref{sect:matt-inc}, we begin by investigating the phenomenology of a
cosmological model containing a second component $X$ with negative
pressure (in the event the energy density is positive), in addition to
a cosmological constant. Since our `missing matter' model (for which
$w_X=-\frac{2}{3}$) is just a special case (albeit a very important
one) of our more generic (but less theoretically well-motivated)
`double dark energy' model (for which $w_X$ is allowed to vary), we
will focus here on the former as being a representative example of the
latter. 

In our analysis, we do not restrict the energy density $\Omega_X$ (at
any epoch) to be positive. Although once widely accepted, the trace,
strong, null, weak and dominant energy conditions all now have a
somewhat weakened status following recent evidence of violations in
physical systems ranging from neutron stars to inflationary cosmology,
and in particular from the physics of scalar
fields~\cite{Barcelo02,Curiel17}.  Given this ongoing historical
revision of the energy conditions, it seems appropriate to continue in
the tradition of letting the observational data take precedence over
theoretical prejudice. Indeed, from a Bayesian perspective, it seems
prudent not to impose a prior that assigns zero probability
density to negative values of $\Omega_X$, since this may exclude
outcomes that are implied by the data.

The effect of the additional component $X$ on the global expansion
history of the universe depends only on the equation-of-state
parameter $w_X$, whereas its effect on the evolution of perturbations
will also depend on the nature of the component $X$, in particular its
assumed dynamical properties. We therefore consider these two issues
separately.

\subsection{Background evolution}

The global expansion history of the cosmological model is most
conveniently represented through the distance-redshift
relation. Indeed, comparing the predicted relation between the
luminosity distance $d_L$ and redshift $z$ of an object with
observations of astronomical `standard candles', such as Type-Ia
supernovae, has provided the most direct and convincing evidence that
the expansion of the universe is accelerating.

\begin{figure}[tbp]
\centering
\includegraphics[width=0.6\textwidth]{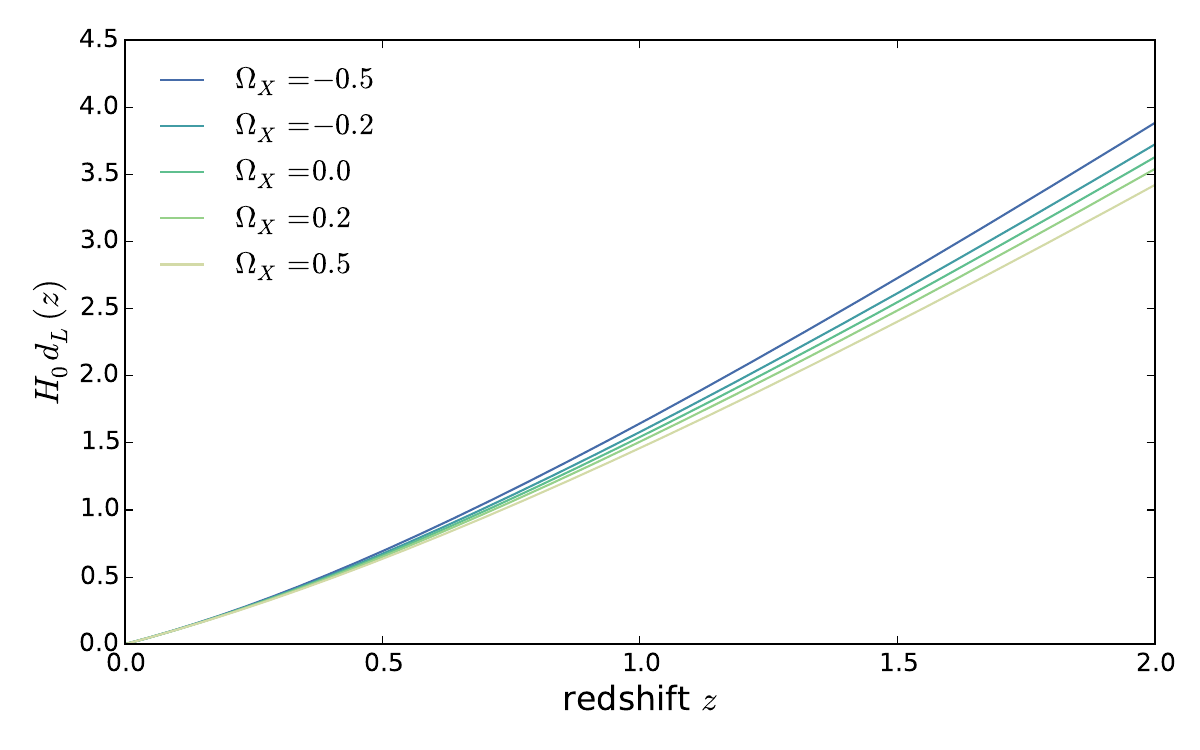}
\caption{Dimensionless luminosity distance $H_0d_L(z)$ as a function
  of redshift $z$ for a concordance $\Lambda$CDM cosmology with an
  additional component $X$ with equation-of-state parameter
  $w_X=-\frac{2}{3}$, for different values of $\Omega_{X,0}$ (and adjusted $\Omega_{\Lambda,0}$).}
\label{fig:d_L}
\end{figure}

The luminosity distance to an object at redshift $z$ is given by
\begin{equation}
d_L(z) = (1+z)\frac{S_k(\sqrt{|\Omega_{k,0}|}\, \chi(z))}{\sqrt{|\Omega_{k,0}|}},
\end{equation}
where $S_k(x)=\sinh x$, $x$, $\sin x$ for spatial curvature parameter
$k=-1$, $0$, $+1$ respectively, and the comoving radial coordinate
$\chi(z)$ is determined by the expansion history:
\begin{equation}
\chi(z)= \int_0^z \frac{d\bar{z}}{H(\bar{z})},
\end{equation}
where $H(z)$ is obtained from the Friedmann equation (\ref{eq:Fried}).
The inclusion of the $\Omega_{X,0}$ into (\ref{eq:Fried}) thus
directly affects the expansion history embodied in $H(z)$, and hence
can serve either to increase or decrease the luminosity distance
$d_L(z)$ to an object at redshift $z$. Fig.~\ref{fig:d_L} illustrates
this effect for a few representative values of $\Omega_{X,0}$.  If
$\Omega_{X,0}>0$, the apparent luminosity is increased and hence the
luminosity distance is reduced compared to the standard $\Lambda$CDM
model. The opposite effect occurs for $\Omega_{X,0}<0$.

The power of the luminosity distance as a cosmological probe resides
in the fact that it can be simply related to apparent brightness
$m(z)$ obtained directly from a set of standard candles, each (assumed
to be) of absolute magnitude $M$, namely
\begin{equation}
m(z) =  M +  5\log_{10}\left[ \frac{d_L(z)}{\rm Mpc} \right] +25,
\end{equation}
where the constant offset ensures the usual convention that $m=M$ for
an object at $d_L = 10$~pc. Type-Ia supernovae constitute a set of
`standardizable candles' that can be used to constrain cosmological
models in this way \cite{SCP}.

It should be pointed out that, for the background evolution, the
combination of a cosmological constant with $w_\Lambda=-1$ and an
additional component $X$ with constant $w_X$ is equivalent, under
certain conditions outlined below, to a single dark energy component
with a time-varying equation-of-state parameter $w_{\rm eff}(a)$ given
by the ratio of the combined pressure of the two components to their
combined density \cite{Gong07}, namely
%
\begin{equation}
w_{\rm eff}(a) =
\frac{-\Omega_{\Lambda,0}+w_X\Omega_{X,0}a^{-1}}
{\Omega_{\Lambda,0}+\Omega_{X,0}\,a^{-1}}.
\label{eq:weff}
\end{equation}
Examples of such models have been studied extensively
\cite{CPL1,JBP,Sendra12,Rubin09,Akarsu2015}, albeit not with the
particular form of $w_{\rm eff}(a)$ given above.  It is clear that the
variation of $w_{\rm eff}$ with either $a$ or redshift $z$ is
non-linear, so $w_{\rm eff}(a)$ is not contained within either of the
common $w(z) =w_0 + w_1z$ or $w(a)=w_0+w_a(1-a)$ parameterisations.
More importantly, it should be noted that if $\Omega_{\Lambda,0}$ and
$\Omega_{X,0}$ have different signs, as we allow in our analysis in
Section~\ref{sec:analysis}, then $w_{\rm eff}(a)$ becomes singular at
$a=|\Omega_{X,0}/\Omega_{\Lambda,0}|$. Thus, if $\Omega_{\Lambda,0}$
or $\Omega_{X,0}$ (or both) are allowed to take positive and negative
values, then our missing matter (or double dark-energy) model is {\it
  not}, in general, described by a single time-varying dark-energy
component. Nonetheless, it is worth comparing the evolution of $w_{\rm
  eff}$ with $a$ or $z$ implied by (\ref{eq:weff}) with current
constraints for a single time-varying dark-energy component. 
\begin{figure}[tbp]
\centering
\includegraphics[width=0.6\textwidth]{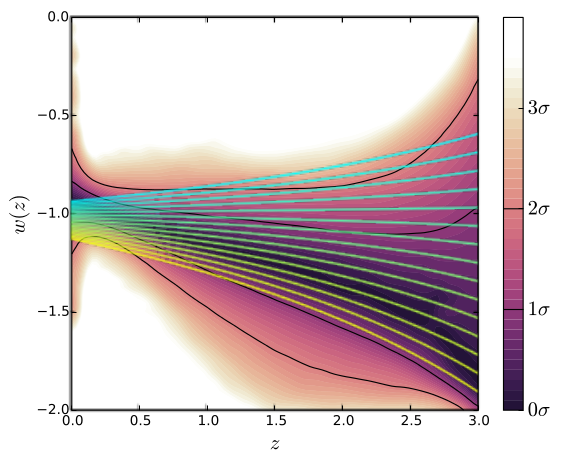}
\caption{The evolution of $w_{\rm eff}$ in equation (\ref{eq:weff})
  with redshift $z$ (solid coloured curves) for $\Omega_{\Lambda,0} =
  0.8$, $\Omega_{X,0} =-0.1$, and $w_X$ ranging from $-1.4$ (blue
  line) to $0$ (yellow line) in steps of $0.1$. These curves are
  overlaid on a `free-form' reconstruction of $w(z)$ for a single
  time-varying dark energy model using Planck 2015 era CMB, BAO, SNIa
  and Lyman-$\alpha$ data, reproduced from \cite{Hee17}, which shows
  the posterior probability $\Pr(w|z)$, with colour scale in
  confidence interval values, and the $1\sigma$ and $2\sigma$
  confidence intervals plotted as black lines.}
\label{fig:weff}
\end{figure}
Such a comparion is plotted in Fig.~\ref{fig:weff}, where we have
assumed the values $\Omega_{\Lambda,0} = 0.8$, $\Omega_{X,0} =-0.1$ in
(\ref{eq:weff}), which are consistent with those obtained in
Section~\ref{sec:results} from our analysis of observational data, and
$w_X$ ranges from $-1.4$ (blue line) to $0$ (yellow line) in steps of
$0.1$. It is clear that the resulting $w_{\rm eff}(z)$ curves are
indeed consistent with the constraints on $w(z)$ for a single
time-varying dark energy model. For the assumed values of
$\Omega_{\Lambda,0}$ and $\Omega_{X,0}$, it is worth noting that
$w_{\rm eff}$ in (\ref{eq:weff}) becomes singular at $a=1/8$, or
equivalently $z=7$, and hence corresponds closely to a single
time-varying dark energy model over the range of redshifts for which
observational constraints are available.

\subsection{Evolution of perturbations}
\label{sec:pert}

An additional component $X$ will affect the growth of perturbations
through its contribution to $H(z)$ and the evolution of the matter
density. Moreover, we assume here that $X$ has the same dynamical
behaviour as that usually assumed for a generic dark energy component.
In particular, we use the CAMB \cite{CAMB} dark-energy module developed
by \cite{PPF}, in which dark energy is assumed itself to exhibit
Gaussian adiabatic perturbations. It is worth noting that, as the
equation-of-state parameter approaches $-1$, the effects of the dark
energy perturbations disappear, as one would expect for a pure
cosmological constant.\footnote{It should be borne in mind, however,
  that a possible physical instantiation of an additional component
  $X$ with $w_X=-\frac{2}{3}$ could be in the form of domain-wall
  topological defects, for example, in which case the effect on the
  generation and evolution of perturbations may be very different to
  that assumed here.} We modified the CAMB software to include our
additional component and calculate the predicted power spectra of
cosmic microwave background (CMB) anisotropies and matter
perturbations, for several values of $\Omega_{\rm X,0}$;
the values of the remaining cosmological parameters were set to their
standard concordance values with $\Omega_{\Lambda,0}$ varying accordingly to ensure that $\sum_{i{=}r, m, k, X, \Lambda} \Omega_i = 1$.

\begin{figure}[tbp]
\centering
\includegraphics[width = 0.6\textwidth]{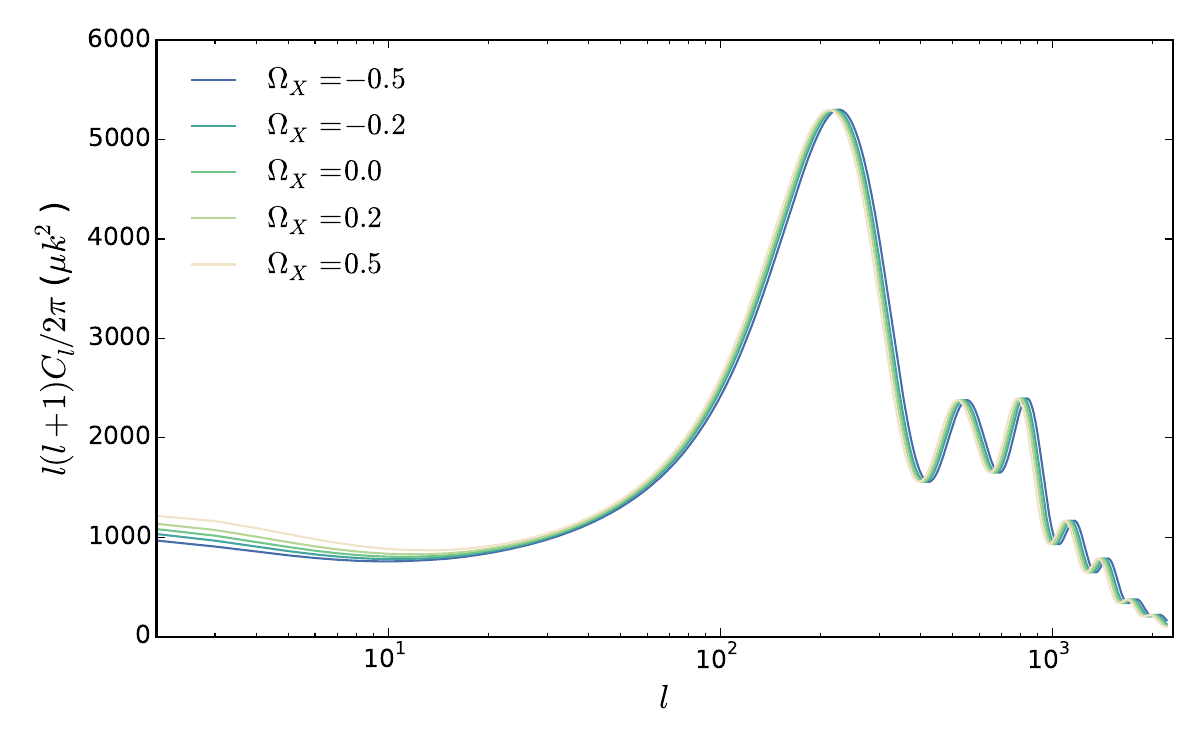}
\caption{CMB power spectra for a concordance $\Lambda$CDM model with
  an additional component $X$, with equation-of-state parameter
  $w_X=-\frac{2}{3}$, for several values of $\Omega_{X,0}$.}
\label{fig:CMB}
\end{figure}

We plot the CMB power spectra in Fig.~\ref{fig:CMB}, from which we see
that, as one might expect, the main effect of a non-zero $\Omega_{\rm
  X,0}$ is to shift the positions of the acoustic peaks, which are
sensitive to the spatial geometry of the universe, and hence depend on
the total energy density of all the components.  Thus, one would
expect constraints on $\Omega_{\rm X,0}$ from CMB observations to be
tightly correlated with the constraints on $\Omega_{\Lambda}$ and
$\Omega_k$. For positive values of $\Omega_{\rm X,0}$, we also see an
 enhancement of power on the largest scales from the
late-time ISW effect. The CMB power spectrum is now well-constrained
by observations over a wide range of scales.

\begin{figure}[tbp]
\centering
\includegraphics[width = 0.6\textwidth]{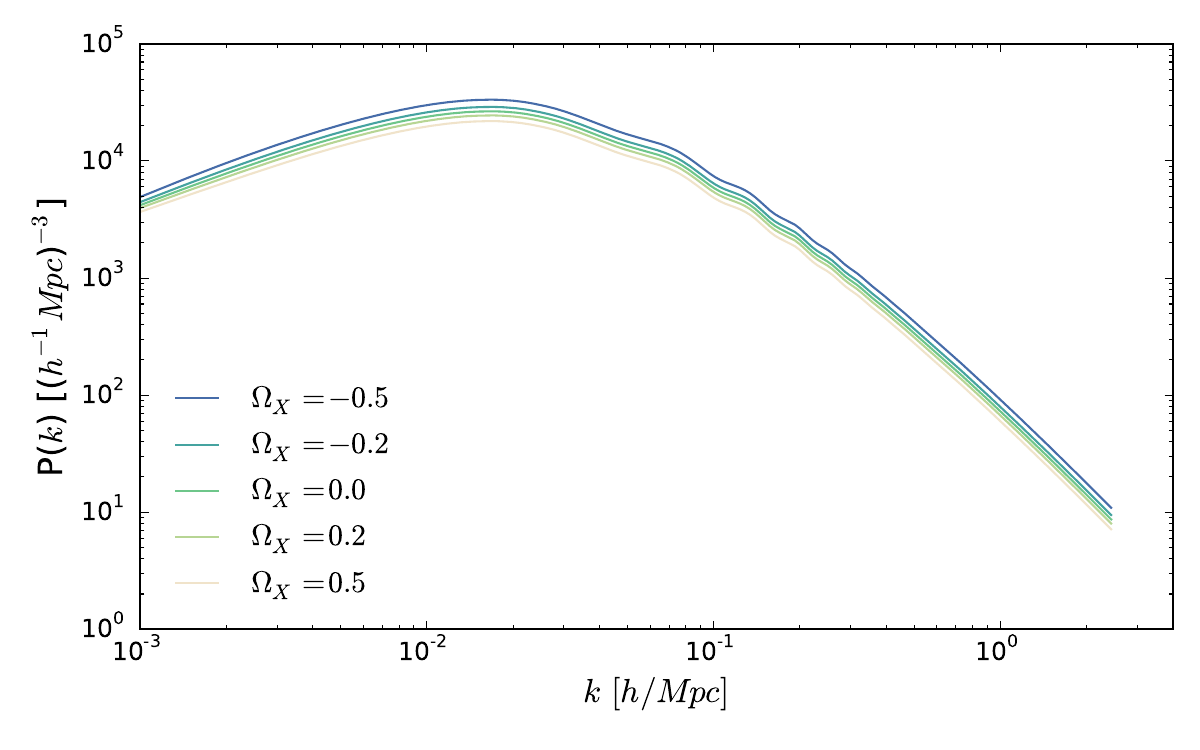}
\caption{Matter power spectra for a concordance $\Lambda$CDM model with
  an additional component $X$, with equation-of-state parameter
  $w_X=-\frac{2}{3}$, for several values of $\Omega_{X,0}$.}
\label{fig:MPk}
\end{figure}

In Fig.~\ref{fig:MPk}, we plot the predicted matter power spectra for
different values of $\Omega_{X,0}$; again the other parameters are set
to their concordance values, with $\Omega_{\Lambda,0}$ varied to incorporate the missing matter density. We see that the dominant effect of the
additional component is on the normalisation of the matter power
spectrum. The amplitude of fluctuations is supressed for $\Omega_{X,0}
>0$ and enhanced for $\Omega_{X,0}<0$. By contrast, the positions of
the acoustic oscillations, which depend on the matter density,
are unaffected by the introduction of the additional component.

\begin{figure}[tbp]
\centering
  \includegraphics[width = 0.6\textwidth]{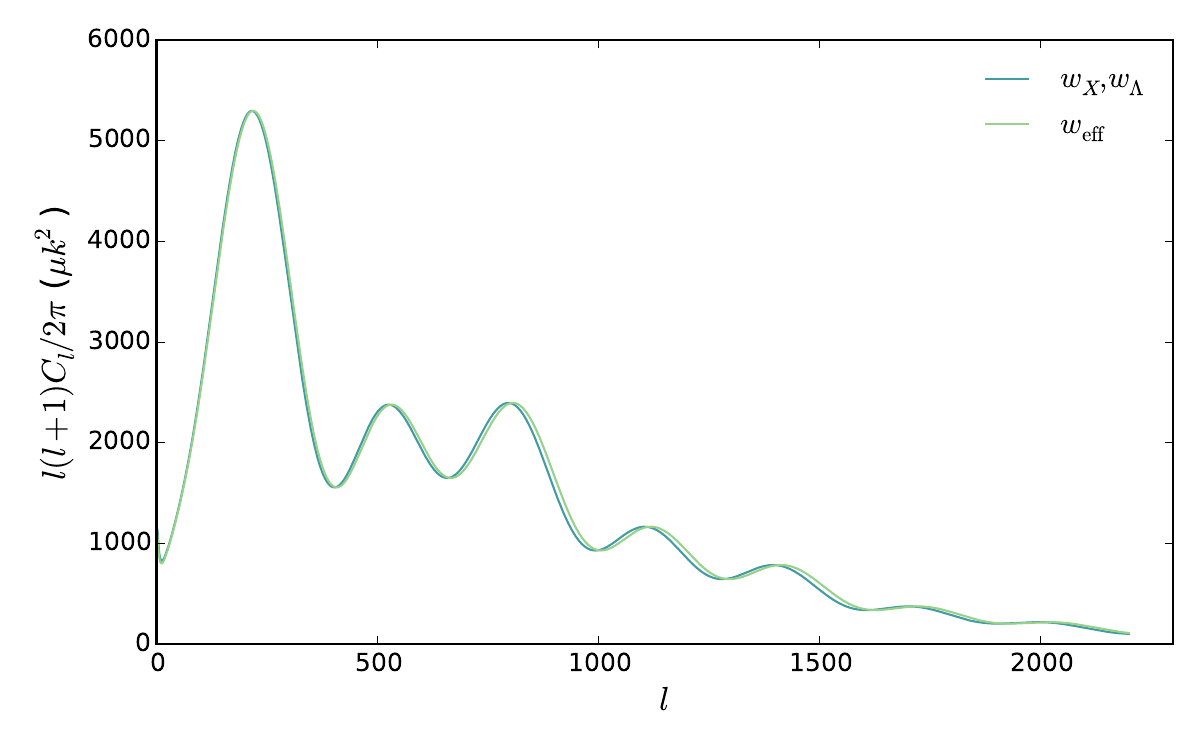}
\hfill
  \includegraphics[width = 0.6\textwidth]{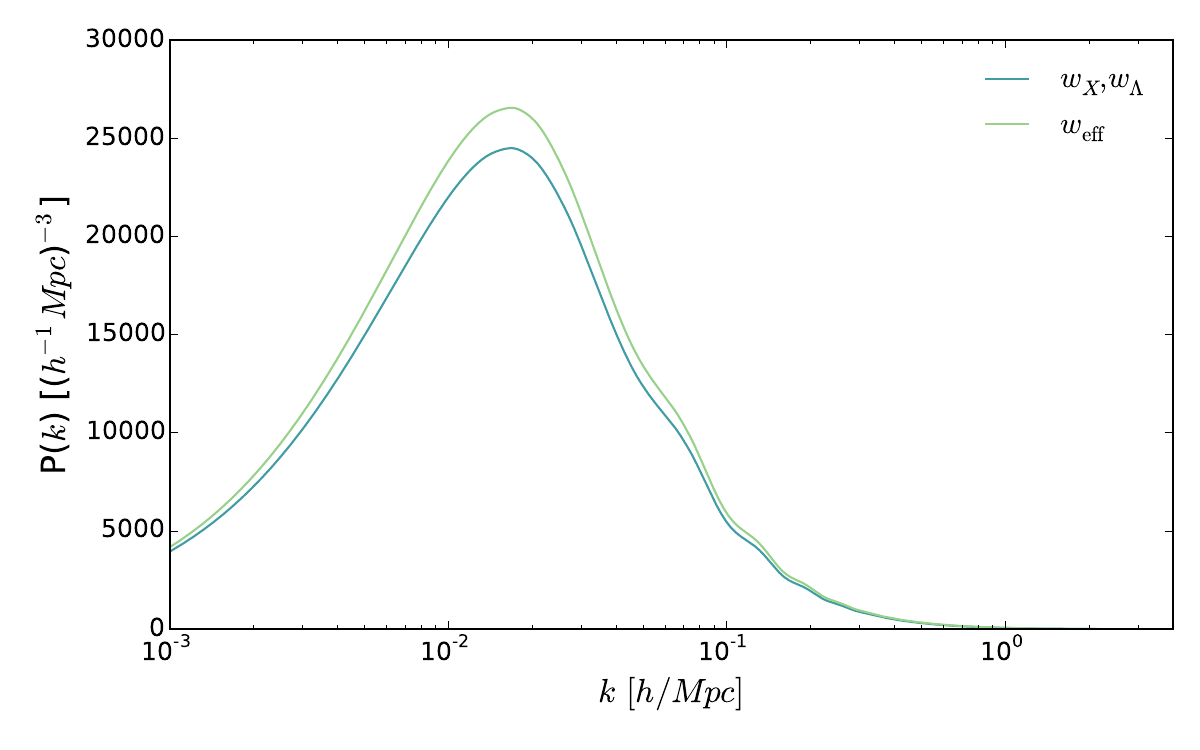}
\caption{CMB power spectra (top) and matter power spectra (bottom)
  for: a concordance $\Lambda$CDM model with an additional component
  $X$, with equation-of-state parameter $w_X=-\frac{2}{3}$ and density
  $\Omega_{X,0}=0.2$ (dark blue line); and a CDM model with a single
  time-varying dark energy component with effective equation-of-state
  parameter $w_{\rm eff}(a)$ defined in (\ref{eq:weff}) (light green line).
\label{fig:weffperts}}
\end{figure}

It is worth noting that, although the background evolution of the
universe is identical for our missing matter (or double dark-energy)
model and for a model with a single time-varying dark energy component
defined by (\ref{eq:weff}) (provided $\Omega_{\Lambda,0}$ and
$\Omega_{X,0}$ have the same sign), the evolution of perturbations is,
in general, {\it different} for the two cases.  This is true even in
the simplest case where one assumes the same dynamical behaviour for
the generic dark energy components in the two models, namely that they
exhibit Gaussian adiabatic perturbations.  This is illustrated in
Fig.~\ref{fig:weffperts}, in which we plot the CMB and matter power
spectra for a specific example of each model.
Consequently, we reiterate our earlier comment that the many previous
studies of models containing a single time-varying dark-energy
component are {\it not} equivalent to the study presented here.


\section{Analysis}
\label{sec:analysis}

\noindent
We now perform a Bayesian parameter estimation and model comparison
analysis of our `missing matter' and `double dark energy' models,
using recent cosmological observations. In particular, we use the
Planck 2015 data release temperature measurements~\cite{PlanckLikelihood}
and lensing data~\cite{PlanckLensing}. In addition to CMB data, we include distance
measurements of 740 Supernovae Ia from the SNLS-SDSS collaborative effort
called the joint light-curve analysis (JLA; \cite{JLA}) and several Baryon Acoustic
Oscillation (BAO; \cite{BAODR11, BAO6DF, BAOMGS, BAOBusca, BAOAndreu})
measurements of distance.

Throughout the analysis we consider purely Gaussian adiabatic scalar
perturbations and neglect tensor contributions.  We assume a modified
$\Lambda$CDM model specified by the following parameters: the physical
baryon density $\Omega_{\rm b} h^2$ and CDM density $\Omega_{\rm DM}
h^2$, where $h$ is the dimensionless Hubble parameter such that
$H_0=100h$ kms$^{-1}$Mpc$^{-1}$; the curvature density $\Omega_{k,0}$
of the universe; $\theta$, which is $100\times$ the ratio of the sound
horizon to angular diameter distance at last scattering surface; the
optical depth $\tau$ at reionisation; and the amplitude $A_{\rm s}$
and spectral index $n_{\rm s}$ of the primordial perturbation spectrum
measured at the pivot scale $k_0=0.05$ Mpc$^{-1}$. We also include
17 nuisance parameters associated with the Planck and JLA datasets.
The ranges of the uniform priors assumed on the standard $\Lambda$CDM
parameters are listed in Table~\ref{tab:priors}, with nuisance parameter
priors set to the advised values. Our hypothetical
additional component is characterised by its density parameter
$\Omega_{X,0}$ and equation-of-state parameter $w_X$. We assume a
uniform prior on $\Omega_{X,0}$ in the range $[-1,2]$ throughout. For
the missing energy model, we have $w_X=-\frac{2}{3}$, and for the
double dark energy model we assume the uniform prior $w_X =
[-\frac{3}{2},-\frac{1}{2}]$.

To carry out the exploration of the parameter space, we first
incorporate the extra component into the standard cosmological
equations, by performing the minor modifications to the {\sc CAMB} code
\cite{CAMB} described in Section~\ref{sec:pert} (which implement a parameterised post-Friedmann (PPF)
prescription for the dark energy perturbations \cite{PPF}).  We then
include into the {\sc CosmoMC} software \cite{Cosmo} a
fully-parallelised version of the nested sampling algorithm {\sc
  PolyChord} \cite{PolyChord2015a,PolyChord2015b}, which significantly increases the
efficiency of calculating the Bayesian evidence and also reliably
produces posterior samples even from distributions with multiple modes
and/or high dimensionality.
A suitable guideline for making qualitative conclusions has been laid
out by Jeffreys~\cite{Jeffreys}: if $\mathcal{B}_{ij}< 1$ model $i$
should not be favoured over model $j$, $1<\mathcal{B}_{ij}<2.5$
constitutes significant evidence, $2.5<\mathcal{B}_{ij}<5$ is strong evidence, while
$\mathcal{B}_{ij}>5$ would be considered decisive.
\begin{table}
\begin {center}
\begin{tabular}{ll}
\hline
\vspace{0.1cm}
Parameter & Prior range\\
\hline
\vspace{0.1cm}	
$\Omega_{\rm b,0}h^2$ \hspace{1.5cm}    & $[0.019, 0.025]$ \\
$\Omega_{\rm dm,0}h^2$                  & $[0.095, 0.145]$ \\
$\Omega_{k,0}$                          & $[-0.05, 0.05 ]$ \\
$\theta$                                & $[1.03 , 1.05 ]$ \\
$\tau$                                  & $[0.01 , 0.4  ]$ \\
$n_{\rm s}$                             & $[0.9  , 1.1  ]$ \\
$\ln[10^{10} A_{\rm s}]$                & $[2.7  , 4.0  ]$ \\
\hline
\end{tabular}
\caption{Ranges of the uniform priors assumed on the standard
  $\Lambda$CDM parameters in the Bayesian analysis.}
\label{tab:priors}
\end{center}
\end{table}
%


\section{Results}
\label{sec:results}

For comparison purposes, we first assume no additional component $X$,
in order to determine the constraints imposed by the current combined
data sets on the standard $\Lambda$CDM model. In particular, we find
the data indicate the dominance of dark energy in the form of a
cosmological constant with $\Omega_{\Lambda,0}= 0.696 \pm 0.007$,
followed by matter density (dark matter+ baryons) $\Omega_{\rm m,0}
=0.305 \pm 0.007$ , and an almost negligible spatial curvature
$\Omega_{k,0}=-0.0013 \pm 0.0024$.  We also obtain the present Hubble
parameter $H_0=67.78 \pm 0.70$.  The constraints on the other
parameters \{$\theta, \tau, A_{\rm s}, n_{\rm s}$\} remain essentially
unaffected by the introduction below of our additional component $X$,
and so we do not consider them further.

\subsection{Missing matter model}

\begin{figure}[tbp]
\centering
\includegraphics[width=9cm, height=8cm]{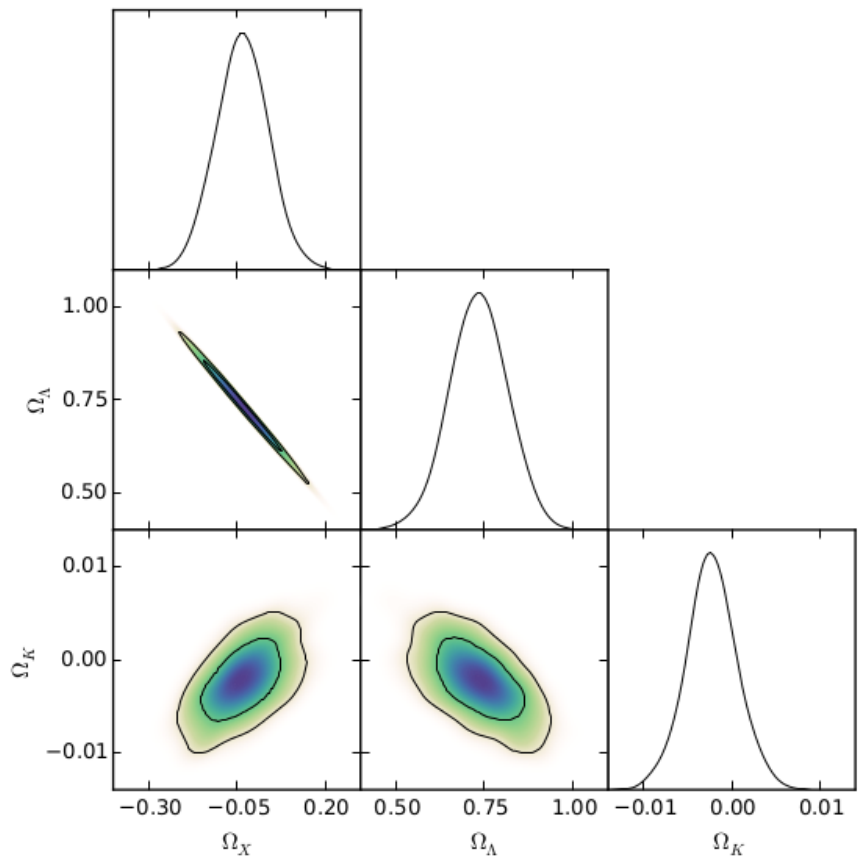}
\caption{1D and 2D marginalised posterior distributions for density
  parameters in the missing matter model (note that $\Omega_{\textrm{m},0} =
  1-\Omega_{\Lambda,0}-\Omega_{k,0}-\Omega_{X,0}$). The 2D constraints
  are plotted with $1\sigma$ and $2\sigma$ confidence contours and the
  cubehelix colour map \cite{Green2011}.}
\label{fig:3CDM}
\end{figure}

The inclusion of a missing matter component $X$ with
$w_X=-\frac{2}{3}$ considerably broadens the parameter constraints.
In particular, we find: $\Omega_{\Lambda,0}= 0.734 \pm 0.083$, which
constitutes an order-of-magnitude increase in the error bars as
compared with the standard $\Lambda$CDM model, $\Omega_{\rm m,0}=
0.302 \pm 0.010$, $\Omega_{k,0}= -0.0023 \pm 0.0029$ and
$H_0=68.10 \pm 1.04$. Figure~\ref{fig:3CDM} shows 1D and 2D
marginalised posterior distributions for the density parameters (note
that $\Omega_{\textrm{m},0} =
1-\Omega_{\Lambda,0}-\Omega_{k,0}-\Omega_{X,0}$). As expected, we
observe a clear degeneracy between $\Omega_{X,0}$ and
$\Omega_{\Lambda,0}$, and slight degeneracy between $\Omega_{X,0}$ and $\Omega_{k,0}$. The 1D constraint on the
density parameter of missing matter is $\Omega_{X,0} =-0.034 \pm 0.075$.
The current data prefer a slightly negative value, which is
difficult to interpret physically, but the errors suggest this not to be a significant favouring. The 1D marginal shows
moderate relative probability even for $\Omega_{X,0} \sim 0.1$, and so
the presence of an appreciable missing matter component cannot be
ruled out. Our results are, however, still consistent with a standard
$\Lambda$CDM model.

This view is supported by our Bayesian model comparison. We find that
the log-evidence difference (or Bayes factor) between the missing
matter model and the standard $\Lambda$CDM model is
$\mathcal{B}_{\Lambda+X, \Lambda} =-1.12 \pm 0.53$. According to
Jeffreys guideline \cite{Jeffreys,Vazquez12}, the inclusion of the
missing matter component is therefore slightly disfavoured, but almost
indistinguishable, from a model perspective given current cosmological
data.

\subsection{Double dark energy model}

We now allow for the equation-of-state parameter $w_X$ for our
additional component to be a free parameter (albeit still independent
of redshift), for which we assume a uniform prior in the range
$w_X=[-\frac{3}{2},-\frac{1}{2}]$. We thus allow for the possibility
that this second dark-energy component could be a form of phantom
energy with $w_X < -1$ \cite{Vazquez12b}.  It should also be pointed
out, however, that this parameterisation for the additional component
necessarily includes a cosmological constant as the special case
$w_X=-1$. This therefore leads to an unavoidable degeneracy between
the additional component and the cosmological constant, and this
should be borme in mind when interpreting the parameter constraints
derived from the cosmological data.

\begin{figure}[tbp]
\centering
\includegraphics[width=10cm, height=8.5cm]{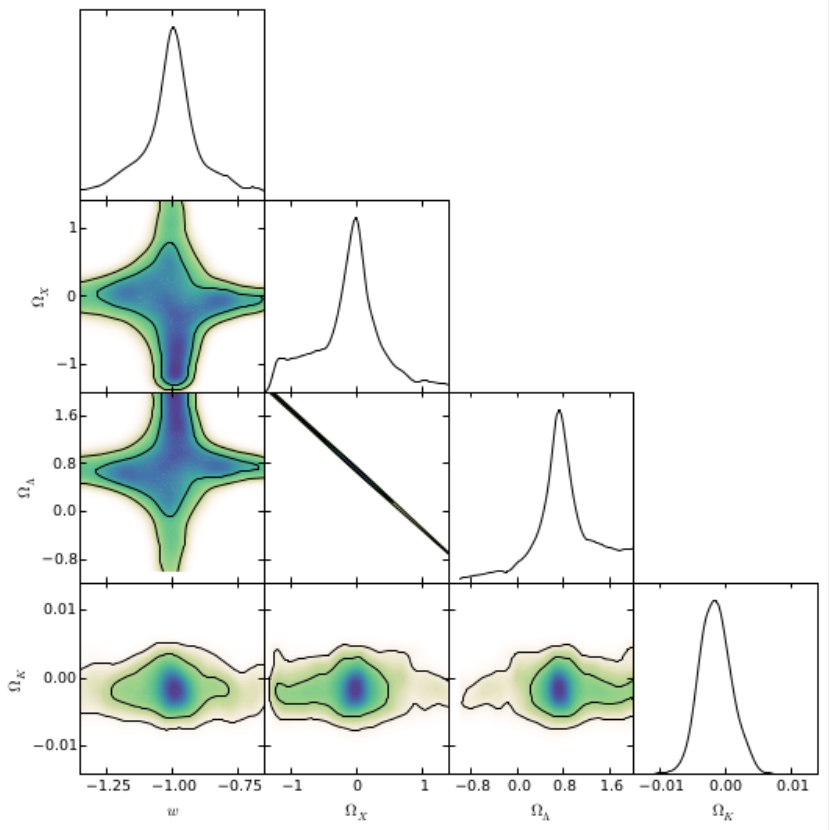}
\llap{\raisebox{5cm}{
  \includegraphics[width=4.5cm, height=4cm]{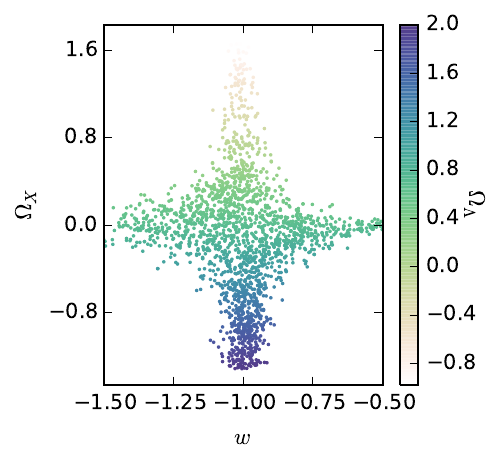}
}}
\caption{1D and 2D marginalised posterior distributions for density parameters
in the double dark energy model (note that $\Omega_{m,0} =
1-\Omega_{\Lambda,0}-\Omega_{k,0}-\Omega_{X,0}$). The 2D constraints
are plotted with $1\sigma$ and $2\sigma$ confidence contours. The
top-right panel shows the 3D posterior distribution in the
$(w_X,\Omega_{X,0},\Omega_{\Lambda,0})$ subspace, where the colour code
indicates the value of $\Omega_{\Lambda,0}$ using the cubehelix colour map \cite{Green2011}.
\label{fig:XCDM}}
\end{figure}

Figure~\ref{fig:XCDM} shows the resulting 1D
and 2D marginalised posterior distributions for $w_X$ and the density
parameters in the model (once again, note that $\Omega_{m,0} =
1-\Omega_{\Lambda,0}-\Omega_{k,0}-\Omega_{X,0}$). At the top-right of
the figure we also give a representation of the 3D posterior in the
$(w_X,\Omega_{X,0},\Omega_{\Lambda,0})$ subspace, where the colour
indicates the value of $\Omega_{\Lambda,0}$.

The 1D constraints on the standard parameters are as follows:
$\Omega_{\Lambda,0}= 0.797 \pm 0.556$, $\Omega_{\rm m,0}= 0.305 \pm
0.009$, $\Omega_{k,0}=-0.0015 \pm 0.0024$, $H_0=67.86 \pm 1.01$. The
constraints on the parameters describing the additional second
dark-energy component may be given as $w_X=-1.01 \pm 0.16$ and
$\Omega_{X,0}= -0.101 \pm 0.557$, although these numbers obscure the
nature of the marginal $\left(w_X,\Omega_{X,0}\right)$-space and
$\left(w_X,\Omega_{\Lambda,0}\right)$-space distributions slightly.
These results are clearly consistent with a standard $\Lambda$CDM
model, although the inclusion of the additional dark-energy component
has again resulted in the uncertainties in the constraints on the
standard parameters being much larger than those obtained assuming a
$\Lambda$CDM model. Indeed, the 1D marginal for $\Omega_{X,0}$ shows
moderate relative probability even for $\Omega_{X,0} \sim \pm 0.3$,
although this is likely due to a value of $w{=}-1$ simply reproducing
the $\Lambda$CDM model.

Moreover, the 2D and 3D marginal distributions in Fig.~\ref{fig:XCDM}
have interesting features that are worth
noting. As might be expected, we again see a pronounced
degeneracy between $\Omega_{\Lambda,0}$ and $\Omega_{X,0}$.
The marginal distribution in $(\Omega_{X,0},\Omega_{\Lambda,0})$
subspace shows a strong correlation between these energy densities
that would imply the potential for a trade-off between them. One might
be concerned, however, that the marginal distribution plotted is
strongly dominated by the contribution (after marginalising over
$w_X$) from near $w_X = -1$. If so, one could then not infer the
potential of a trade-off between these two energy densities at (any)
other values of $w_X$. To investigate this possibility, we also
calculated the conditional distributions in
$(\Omega_{X,0},\Omega_{\Lambda,0})$ subspace for a small set of fixed
$w_X$-values in the range $[-0.7,-1.3]$. The resulting distributions
were, however, very similar to that plotted in figure~\ref{fig:XCDM},
and so indicating that the two energy densities can indeed be
traded-off against one another.

Also of interest is our Bayesian model comparison, which finds that
the log-evidence difference (Bayes factor) between the double dark
energy model and standard $\Lambda$CDM is $\mathcal{B}_{\Lambda+X,\Lambda}
= -0.43 \pm 0.45$. This shows that neither model is preferred over the
other with any significance; indeed they are in the indistinguishable
range of Jeffreys guideline and identical within $1\sigma$ of the error
on the evidence calculation. Thus, the two additional parameters
$\Omega_{X,0}$ and $w_X$ in the double dark energy model allow it the
freedom to fit the data sufficiently better than $\Lambda$CDM to
compensate for the corresponding increase in the prior volume, and
hence the model is not penalised by the evidence. The Bayes factor stated
is likely also aided by the broadening of posteriors on some of the parameters,
as this implies a lower Occam factor associated with those parameters.


\section{Discussion and Conclusions}
\label{sec:conclusions}

We have investigated the possibility that there exist two dark-energy
components in the universe: a cosmological constant, with $w=-1$; and
an additional component $X$ with equation-of-state parameter $w_X$. In
the first instance, we fix the equation-of-state parameter of $X$ to
the value $w_X=-\frac{2}{3}$. Assuming the canonical values for
equation-of-state parameters of the other components, this `missing
matter' model corresponds to the special case in which the additional
component is \textit{required} for the Friedmann equation written in
terms of conformal time $\eta$ to be form invariant under the
reciprocity transformation $\tilde{a}(\eta) =\alpha^2/a(\eta)$, where
$\alpha^2$ is a constant, which is relevant to scenarios such as
Penrose's conformal cyclic cosmology (CCC) proposal. Foregoing this
requirement, we then consider the more general `double dark energy'
model, in which $w_X$ is a free parameter assumed to have uniform
prior in the range $w_X=[-\frac{3}{2},-\frac{1}{2}]$. For both models,
we perform a Bayesian parameter estimation and model selection
analysis, relative to standard $\Lambda$CDM, using recent cosmological
observations of cosmic microwave background anisotropies, Type-Ia
supernovae and large scale-structure.

For the missing matter model, the introduction of the additional
component $X$ significantly broadens the constraints on the standard
parameters in the $\Lambda$CDM model, but leaves their best-fit values
largely unchanged. The 1D marginalised constraint on the missing
matter density parameter is $\Omega_{X,0} =-0.034 \pm 0.075$.  Thus,
current cosmological observations prefer a slightly negative value,
the interpretation of which is unclear, but the posterior on this
parameter is sufficiently broad that significant relative probability
exits even for $\Omega_{X,0} \sim 0.1$, and so the presence of a
missing matter component cannot be ruled out. To support this
conclusion, our results are consistent with $\Lambda$CDM and our
Bayesian model selection analysis suggests the missing matter model to
be almost indistinguishable from $\Lambda$CDM, with a Bayes factor of
$-1.12 \pm 0.53$ log-units of evidence.

For the double dark energy model, the constraints on standard
$\Lambda$CDM parameters are again considerably broadened. The 1D
marginalised constraints on the vacuum and second dark energy
component are $\Omega_{\Lambda,0}= 0.797 \pm 0.556$ and $\Omega_{X,0}=
-0.101 \pm 0.557$ (with $w_X=-1.01 \pm 0.16$), respectively, which are
again consistent with $\Lambda$CDM. Once more, however, the 1D
marginalised posterior on $\Omega_{X,0}$ is sufficiently broad that
even $\Omega_{X,0} \sim \pm 0.3$ is not ruled out.
We also find that the double dark energy model has a similar Bayesian evidence
to $\Lambda$CDM, and hence neither model is preferred over the other.


\acknowledgments

This work was carried out largely on the Cambridge High Performance Computing
cluster, DARWIN, and the COSMOS Shared Memory computing system at DAMTP\@.
JAV is supported by CONACYT M\'exico. SH is supported by STFC in the
UK\@.





\end{document}